\documentclass{article}
\usepackage{amsmath}

\usepackage{amssymb}

\newtheorem{Theorem}{Theorem}

\newcommand{\Proof}{\noindent {\em Proof:}\ \ }
\newcommand{\QED}
   {\hfill$\hbox{\vrule height1.3ex width1.3ex depth.1ex}\ $
    \par\smallskip}
\newcommand{\bref}[1]{(\ref{#1})}

\newcommand{\half}{\tfrac{1}{2}}

\newcommand{\ric}{\mbox{\sc{R}}}
\newcommand{\cy}{\mbox{\sc{C}}}
\newcommand{\rsc}{\mbox{R}}

\newcommand{\tr}{\mathrm{tr}_g}
\newcommand{\discr}{\mathcal{D}}

\title{\bf  A Note on Static Metrics }

\author{Robert Bartnik\protect{\thanks{%
School of Mathematical Sciences,
Monash University.}}\\ and \\
Paul Tod\thanks{%
Mathematical Institute, Oxford, OX1 3LB.}}
\date{}

\begin{document}
\maketitle

\begin{abstract}\vskip 3mm

  Conditions are given which, subject to a genericity condition on the
  Ricci tensor, are both necessary and sufficient for a 3-metric to
  arise from a static space-time metric.

\vskip 4.5mm



\end{abstract}

\vskip 12mm

The vacuum field equations of General Relativity reduce for a static
solution to a coupled system involving a (Riemannian) spatial metric
$g_{ij}$ and a potential function $V$ equal to the square-root of the
norm of the Killing vector. The system is over-determined and one can
ask whether a given metric allows the existence of a $V$ for which the
equations are satisfied. This is the question: {\em{when is a 3-metric
    static?}} which in turn is the simplest case of a larger question:
{\em{when are Cauchy data for a vacuum solution actually data for a
    static solution?}}

In this note we give necessary and sufficient conditions on a 3-metric
for it to be static, in the case when the Ricci tensor viewed as an
endomorphism of vectors has distinct eigenvalues, which is the generic
case. In this case, our method gives an explicit candidate for the
gradient $dV$ in terms of the curvature. If the Ricci tensor is
degenerate, in the sense of having a repeated eigenvalue, the method
fails. One can then ask if it is possible to have two different
potentials with the same metric, a problem solved in \cite{tod}. Our
methods can also be used to give information on a number of related
geometrical equations, which we shall note below. The larger question
above can also be solved, and will be considered elsewhere.

In terms of a metric $g_{ij}$ with Ricci tensor $R_{ij}$ and a
potential $V$, the static vacuum field equations are as follows:
\begin{eqnarray}
 \label{s1}
  \ric_{ij} &=&V^{-1} \nabla_i\nabla_jV
\\
\label{s2}
\Delta_g V &=&0,
 \end{eqnarray}
 where $\Delta_g$ is the Laplacian for $g$ (for these equations see, for example, \cite{KSHM}). It follows from (\ref{s1})
 and (\ref{s2}) that the scalar curvature vanishes. Define the
 Cotton-York tensor, as usual, by
\begin{equation} \label{cy}
   \cy_{ij} =  \epsilon_{j}^{\
   pq}(\nabla_q\ric_{ip}-\tfrac{1}{4}g_{ip}\nabla_q\rsc),
 \end{equation}
 then $\cy_{ij}$ is symmetric, trace-free and divergence-free in the
 sense that $\nabla^j\cy_{ij}=0$.  Differentiating \bref{s1} and
 applying \bref{cy} and the Ricci identity appropriate for dimension 3
 leads to
\begin{equation}\label{main}
   V\cy_{ij}  = -\epsilon_{j}^{\
   pq}(2\ric_{ip}\delta^k_{q}+g_{ip}\ric_{q}^{k})\nabla_kV, 
 \end{equation}
 which gives a system of five equations (since both sides are
 symmetric and trace-free) for the three components of $U_i
 :=V^{-1}\nabla_iV$.  For this system to be solvable for $U_i$ two
 linear constraints must hold on the five components of the tensor
 $\cy_{ij}$. These constraints are
\begin{equation}
   \cy^{ij}\ric_{ij}=0,\quad \cy^i_j\ric^j_k\ric^k_i = 0,
 \end{equation}
 which, for brevity, we may write as $\cy\cdot\ric=0$ and
 $\cy\cdot\ric^2=0$. The necessity of these conditions follows readily
 from (\ref{main}); sufficiency can be seen by calculating in the
 Ricci eigenframe. In a frame which diagonalises the Ricci tensor, say
 $ \ric = \mathrm{diag}(\lambda,\mu,\nu) $ with $\lambda+\mu+\nu=0$,
 we see from (\ref{main}) that $\cy$ is purely off-diagonal,
 \begin{equation}
   \cy \simeq \left[
   \begin{array}{ccc} 0&c&b \\ c&0&a \\ b&a&0    \end{array} \right]
\end{equation}
and for $U$, provided $\lambda$, $\mu$ and $\nu$ are distinct, we
obtain the expression
\begin{equation} \label{U1}
   U = [ a/(\mu-\nu), b/(\nu-\lambda), c/(\lambda-\mu) ].
 \end{equation}
 Thus, provided the eigenvalues of the Ricci tensor are distinct, we
 have an explicit expression for the gradient of the (logarithm of
 the) static potential. All that remains is to see whether (\ref{s1})
 and (\ref{s2}) are satisfied, which we do below.

Note that the eigenvalues of the Ricci tensor are 
distinct exactly when the \emph{discriminant}  $\discr$ satisfies
\begin{equation}
   \discr := 4\sigma_2^3 + 27\sigma_3^2 <0,
 \end{equation}
 where $\sigma_1=\lambda+\mu+\nu=\tr \ric=0$, $\sigma_2 = -
 \half{|\ric|^2} = \lambda\mu+\mu\nu+\nu\lambda <0$ and $\sigma_3 =
 \det \ric=\lambda\mu\nu$ are the elementary symmetric functions of
 the eigenvalues of $\ric$.  Note the alternative expressions
 \begin{eqnarray*}
\nonumber
   \discr  &=& -\half |\ric|^6 + 3(\tr \ric^3)^2
\\ 
\nonumber
   &=& -\half (\ric^i_j\ric^j_i)^3 + 3(\ric^i_j\ric^j_k\ric^k_i)^2
\\
   &=& -(\lambda-\mu)^2(\mu-\nu)^2(\nu-\lambda)^2.
\label{D2}
 \end{eqnarray*}
 After some calculation, we may obtain the general form of \bref{U1} in a
 matrix notation as
 \begin{eqnarray}
\nonumber
   U &=& \discr^{-1} (\sigma_2 I+3\ric^2)^2 [\cy,\ric]^*
\\  &=& \tfrac{1}{4} \discr^{-1} (|\ric|^2 I -6\ric^2)^2 [\cy,\ric]^*,
\label{U2}
\end{eqnarray}
where $|\ric|^2:=\ric^i_j\ric^j_i$ and 
$[\cy,\ric]^*:= \epsilon_{ijk}\cy^j_p\ric^{pk}$.  More explicitly,
\begin{equation}
\label{U3}
   U_i = \tfrac{1}{4}\discr^{-1} (|\ric|^2\delta^j_i-6\ric^p_i\ric^j_p)
(|\ric|^2\delta^k_j-6\ric^q_j\ric^k_q) \epsilon_{krs}\cy^r_t\ric^{ts}
\end{equation}

We shall say that a 3-metric $g$ is \emph{Ricci-non-degenerate} if its Ricci
tensor has distinct eigenvalues; equivalently, if $\discr\ne0$.
\begin{Theorem}
  A Ricci-non-degenerate metric $g$, defined on a simply-connected region,
  admits a static potential $V$ exactly when the vector field $U$
  (defined in terms of $g, \ric, \cy$ by \bref{U2} or \bref{U3})
  satisfies
\begin{equation}
\label{Ueq}
   \nabla_iU_j + U_iU_j = \ric_{ij}.
 \end{equation}
\end{Theorem}

\Proof If $g$ is Ricci-non-degenerate and static with potential $V$, then
 $U_i=V^{-1}\nabla_iV$ satisfies
\bref{Ueq} by \bref{s1}, and the above calculations show \bref{U2}
 holds because $\discr<0$.  Conversely, if $U_i$ defined by \bref{U2}
satisfies \bref{Ueq}, then $\nabla_{[i}U_{j]}=0$ and thus $U_i$ is a
gradient, $U_i=\nabla_i\log V$, where the potential $V$ is unique up
to an arbitrary multiplicative constant.  It then follows directly from
\bref{Ueq} that $V$ is a static potential for $g$.  \QED

Equation (\ref{Ueq}) with $U_i$ as in (\ref{U3}) is an equation
directly on the Ricci tensor, and it is of interest to find the form of the derivatives of highest degree. We have the identities
\begin{eqnarray*}
\nabla_i\cy_{jk}&=&\nabla_{(i}\cy_{jk)}+\frac{2}{3}\nabla_{[i}\cy_{j]k}+\frac{2}{3}\nabla_{[i}\cy_{k]j}\\ 
\epsilon^{ijm}\nabla_i\cy_{jk}&=&\Delta \ric_k^m-3\ric^{mn}\ric_{nk}+\delta_k^m\ric^{pq}\ric_{pq}
\end{eqnarray*}
Thus the highest derivatives in (\ref{Ueq}) are $\nabla_{(i}\cy_{jk)}$ and $\Delta\ric_{ij}$. However, given a solution of (\ref{s1}) and (\ref{s2}), $\Delta
\ric_{ij}$ can be expressed as
\[\Delta\ric_{ij}=6\ric_i^k\ric_{jk}-2g_{ij}\ric^{km}\ric_{km}+U^m(\nabla_i\ric_{jm}+\nabla_j\ric_{im}-3\nabla_m\ric_{ij})\]
in terms of $\ric_{ij}$,
$\nabla_i\ric_{jk}$ and $U_i$, and we obtain a necessary condition on just $\nabla_{(i}\cy_{jk)}$ from (\ref{Ueq}) and (\ref{U3}). 

\medskip

\medskip

The conditions found here give an algorithm for testing a metric for
staticity. Assuming Ricci-nondegeneracy, one proceeds by asking the
questions:
\begin{enumerate}
\item
does the scalar curvature vanish?
\item
if so, is $\cy_{ij}$ purely off-diagonal in the Ricci eigenframe?
\item
if so, is $U_i$ from (\ref{U3}) a gradient?
\item
if so, is (\ref{Ueq}) satisfied?
\end{enumerate}
This algorithm can be tested on, for example, spatially homogeneous
3-metrics when it finds all the static cases.

Finally, we note the following two equations which have arisen in the
literature:
\begin{eqnarray}
\nabla_i\nabla_jF&=&\ric_{ij}-\lambda g_{ij}\label{sol}\\
2\nabla_i\nabla_jF&=&F\ric_{ij}.\label{bh}
\end{eqnarray}
The first is the {\emph{gradient Ricci soliton}} equation (see e.g.
\cite{ct}) while the second has arisen in the study of black holes
(see e.g. \cite{reall}). For both equations, the interest is typically
in global solutions on compact manifolds, but our methods in the
3-dimensional case again deduce candidate $dF$ given the metric.

\medskip

\medskip

\noindent{\sc Acknowledgments:} We are grateful to the Isaac Newton
Institute, Cambridge, where part of this work was done, for
hospitality and financial support.  The first author (RAB) also thanks
the Clay Institute for its support.

\end{document}